\documentclass[aps, prd,twocolumn,superscriptaddress,nofootinbib,preprintnumbers]{revtex4-1}
\usepackage{graphicx}
\usepackage{amsmath,amssymb}
\usepackage{hyperref}
\usepackage{braket}
\usepackage{float}
\usepackage[dvipsnames]{xcolor}
\usepackage{soul}
\usepackage{rotating}
\usepackage{multirow}
\usepackage{mathtools}
\usepackage{makecell}

\usepackage[caption = false]{subfig}
\usepackage{txfonts}
\usepackage[margin=0.75in]{geometry}
\usepackage[normalem]{ulem}
\usepackage{graphicx}
\usepackage{subcaption}

\hypersetup{
    colorlinks=true,
    linkcolor=red,
    citecolor=blue,
}

\usepackage{xcolor}

\DeclareMathAlphabet\mathbfcal{OMS}{cmsy}{b}{n}

\newcommand{\beq}{\begin{equation}}
\newcommand{\eeq}{\end{equation}}
\newcommand{\beqa}{\begin{eqnarray}}
\newcommand{\eeqa}{\end{eqnarray}}

\definecolor{darkgreen}{rgb}{0.0, 0.5, 0.0}
\definecolor{darkcyanxf}{RGB}{0.0, 139.0, 139.0}

\newcommand{\redmagic}{\texttt{redMaGiC}\,}
\newcommand{\mice}{\texttt{MICE}\,}

\defcitealias{Battaglia:2012}{B12}
\usepackage{lineno}

\begin{document}
\title[bzev]{Cosmology with imaging galaxy surveys:\\ the impact of evolving galaxy bias and magnification
}

\label{firstpage}

\begin{abstract}
The joint analysis of galaxy clustering and galaxy-shear cross-correlations (galaxy-galaxy lensing) in imaging surveys {constitutes one of the main avenues to obtain cosmological information}. Analyses from Stage III surveys have assumed constant galaxy bias and weak lensing magnification coefficients within redshift bins. We use simple models of bias and magnification evolution to show that plausible levels of evolution  impact cosmological parameter inference significantly for Stage IV surveys and possibly for current datasets as well. The distance weighting factors in galaxy-galaxy lensing cause it to have a different effective mean redshift within a lens bin than galaxy clustering. If the evolution is not included carefully in the theory modeling, this  produces an apparent de-correlation of clustering and lensing even on large  scales and leads to biased cosmological constraints. The level of bias {is sensitive to the redshift and the width} of the lens redshift distributions: for {magnitude-limited-like} samples it is comparable to the statistical precision even for the full DES survey, and exceeds it for Rubin-LSST. We find that ignoring the evolution can lead to approximately $1\sigma$ biases in the cosmological constraints in future Rubin-LSST analyses. We describe how these effects can be mitigated from the modeling and data sides, in particular by using cross-correlations to constrain the evolution parameters. {We also demonstrate that the combination of galaxy clustering and CMB lensing does not suffer from this problem, due to the different distance weighting.}
\end{abstract}

\author{Shivam Pandey}
\affiliation{Columbia Astrophysics Laboratory, Columbia University, 550 West 120th Street, New York, NY 10027, USA}
\affiliation{Department of Physics and Astronomy, University of Pennsylvania, Philadelphia, PA 19104, USA}
\author{Carles S{\'a}nchez} 
\affiliation{Departament de F\'{\i}sica, Universitat Aut\`{o}noma de Barcelona (UAB), 08193 Bellaterra (Barcelona), Spain}
\affiliation{Institut de F\'{\i}sica d'Altes Energies (IFAE), The Barcelona Institute of Science and Technology, Campus UAB, 08193 Bellaterra (Barcelona), Spain}
\affiliation{Department of Physics and Astronomy, University of Pennsylvania, Philadelphia, PA 19104, USA}
\author{Bhuvnesh Jain}
\affiliation{Department of Physics and Astronomy, University of Pennsylvania, Philadelphia, PA 19104, USA}
\author{The LSST Dark Energy Science Collaboration}
\maketitle

\section{Introduction}
The cosmological analysis of the large scale structure (LSS) has entered into a percent level precision era. Multiprobe analyses {in imaging galaxy surveys provide a great avenue to constrain cosmology while also achieving the self-calibration of various sources of systematic uncertainties}. For example, \cite{Abbott2022-xa} jointly analyzed all the six two-point correlations constructed with galaxy positions, galaxy lensing and cosmic microwave background (CMB) lensing. This resulted in self-calibration of multiple systematics parameters and competitive cosmological constraints.

Galaxy biasing, which encapsulates the relation between the clustering of galaxies and dark matter, constitutes an important source of uncertainty for cosmological analyses in galaxy surveys. It is an astrophysical parameter so its value depends on galaxy type and other physical properties (see \cite{Desjacques2016-yp} for a review). Although at large scales it is well described by a linear local bias, this can evolve with redshift as the galaxy sample and physical processes in the Universe also evolve. Within the halo model framework, the large scale bias of galaxies can be written as \cite{Cooray2002-ry}:
\begin{equation}
b_{\rm g}(z) = \frac{1}{\bar{n}_{\rm gal}}\int dz \int dM \, n(M, z) \, N_{\rm g}(M, z) \, b_{\rm halo}(M,z),
\label{eq:bz_HM}
\end{equation}
where $\bar{n}_{\rm gal}(z)$ is the mean density of galaxies at any redshift $z$, $n(M,z)$ is the halo mass function, $N_{\rm g}$ is the total number of galaxies in a halo of mass $M$ and redshift $z$, and $b_{\rm halo}$ is the large scale bias of that halo. We can see that as the halo mass function, the large scale halo bias and the halo occupation distribution (HOD) of the galaxies evolve with redshift, this will directly lead to an evolution of the effective galaxy bias. 

\begin{figure}
\includegraphics[width=0.95\columnwidth]{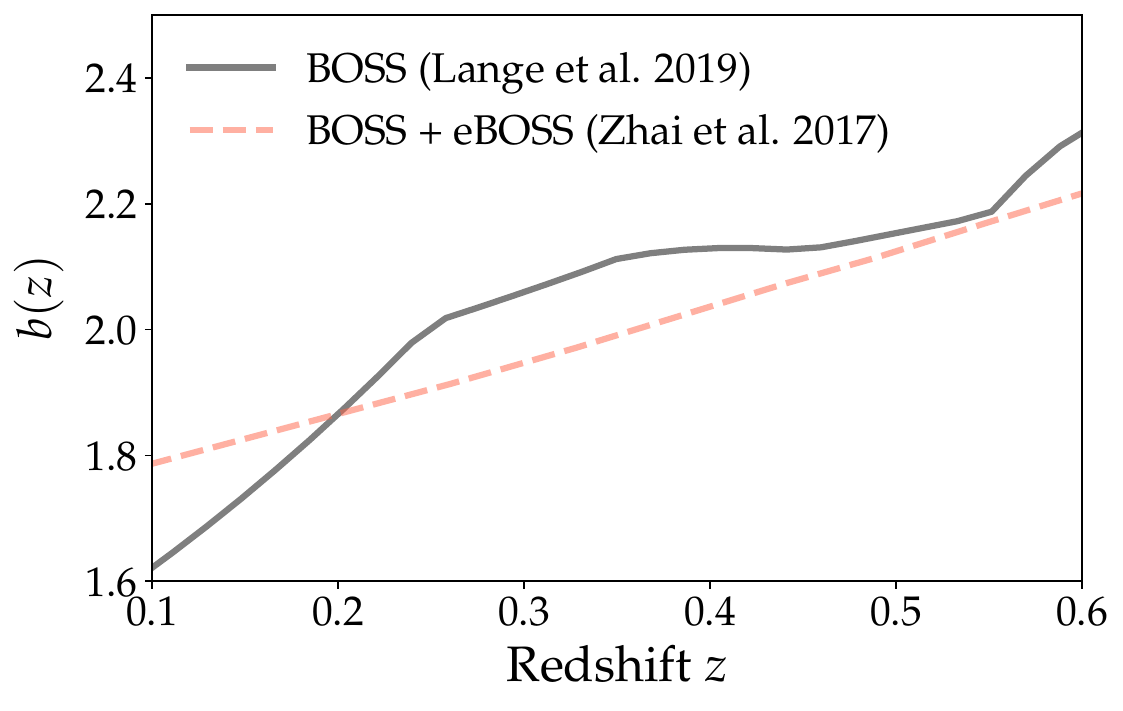}
\caption[]{Comparison of the evolution of the large scale bias using different analysis methods and galaxy samples as described in \cite{Lange2019-fy} and \cite{Zhai2016-tz}. 
}
\label{fig:bz_boss}
\end{figure}

    For example, in \cite{Lange2019-fy}, the galaxies observed in the spectroscopic Sloan Digital Sky Survey (SDSS) survey \citep{Reid2015-vq} were analyzed within the halo model framework to infer the evolution of the total number of galaxies ($N_{\rm g}$) with mass and redshift. We use Eq.~\ref{eq:bz_HM} to infer the evolution of the effective large scales galaxy bias $b_{\rm g}(z)$, as shown in Fig.~\ref{fig:bz_boss}. This analysis particularly used the Baryon Oscillation Spectroscopic Survey (BOSS) galaxy sample. In \cite{Zhai2016-tz}, a similar exploration of galaxy bias evolution is performed while additionally using the extended BOSS (eBOSS) galaxy data \citep{Dawson2015-yk}. We show their bestfit curve obtained when using an evolving HOD model. We see that even for simple galaxy selections, the effective bias can strongly evolve with redshift, showing approximately a 10\% change with 10\% change in redshift (Fig.~\ref{fig:bz_boss}).

In imaging surveys \citep{Sevilla-Noarbe2020-di, Chambers2016-oi, Aihara2019-xy, Dey2018-by, Kuijken2019-fg, Ivezic2019-qc}, the galaxy selection is heavily intertwined with their colors. This can induce steeper evolution in galaxy properties with redshift, and hence causing the galaxy bias to evolve strongly.
Moreover, as the redshift of each galaxy is uncertain, the galaxies are binned into various tomographic bins with a width of $\Delta z_{\rm bin} >~ 0.1$. As we will show in this study, ignoring the evolution of galaxy bias within a tomographic bin can eventually lead to non-trivial impact on large-scale structure analyses using information from galaxy positions and weak gravitational lensing. 

In addition to galaxy bias, other cosmological and astrophysical effects can evolve with redshift as well. One such effect that we explore in this study is lens magnification \citep{Villumsen1995-je}. As the light from galaxies travels  cosmological distances, it gets lensed by the intervening matter. This can change the apparent magnitude of the galaxy and distort the sky area, leading to changes in the number density of observed galaxies. This phenomena, dubbed lens magnification, also depends on the selection of galaxies, such as size cuts \citep{Schmidt2009-eg, Liu2013-ab, 2023MNRAS.520.1541J} or luminosity cuts \citep{LoVerde2007-uz, Moessner1997-np}, which can evolve with redshift. We will show that this evolution, if not accounted for, also leads to biases in cosmological inference. Another such effect would be intrinsic alignments (IA), but in that case the redshift evolution is already considered in state-of-the-art analyses. For instance, in the fiducial DES Y3 analysis \citep{y3-3x2ptkp}, a smooth power-law evolution for IA parameters across source bins is used. For this reason, in this work we focus on galaxy bias and magnification evolution, which has not been studied in detail before. Other more general forms of IA evolution may be worth exploring in further detail as well, but we leave that to future work.

In this paper we will show that characterizing the redshift evolution of galaxy bias and weak lensing magnification is essential for Rubin LSST in its goal to measure the growth of structure at high redshifts. Very little is known observationally about growth at high redshift, and measuring the matter amplitude $\sigma_8(z)$ at percent-level precision or better would provide tight constraints on $\Lambda$CDM, the time-dependence of the dark energy equation of state, and the sum of neutrino masses \citep{2018PhRvD..97l3540S, Sanchez23}.

The paper is organized as follows, in \S~\ref{sec:theory} we describe the theory formulation of the two-point statistics and multiprobe covariance estimation. In \S~\ref{sec:sim_settings} we describe the analysis choices for the simulated tests and in \S~\ref{sec:results} we detail the main findings on the study. We finally conclude in \S~\ref{sec:conclusion}. 


\section{Theory}
\label{sec:theory}
In order to describe the correlations between different tracers, we will follow the notations of \cite{Krause2021-fd} and \cite{Fang2019-jx}. We will focus on the correlations constructed between foreground lens galaxy position, background source galaxy shape and lensing of the CMB. 

\subsection{Two-point statistics}

The transfer function for projected lens galaxy field can be written as:
\begin{equation}
    \Delta_{\rm g, obs}^{i}({\ell}) = \Delta_{\rm g, D}^{i}({\ell}) + \Delta_{\rm g, RSD}^{i}({\ell}) + \Delta_{\rm g, \mu}^{i}({\ell}),
\end{equation}
where, $\Delta_{\rm g, D}^{i}$ is the line-of-sight integration of the lens galaxy overdensity corresponding to tomographic bin $i$. The terms $\Delta_{\rm g, RSD}^{i}$ and $\Delta_{\rm g, \mu}^{i}$ provide the contribution due to the redshift space distortions (RSD) \citep{Kaiser1987-rt, Tanidis2019-cm} and magnification of the lens galaxies \citep{Villumsen1995-je, Tanidis2019-on} respectively. Let us define the terms $T_{\delta}, T_{\theta}$ and $T_{\phi + \psi}$ as the transfer function of matter perturbations, velocity divergences and scalar metric perturbations respectively and are related to each other through cosmological equations of motion. Then the lens galaxy overdensity can be individually explicitly written as \citep{Chisari2019-el} :
\begin{align}
\Delta_{\rm g, D}^{i}({\ell})  = \int dz \, n_{\rm g}^{i}(z) \, b_{\rm g}^{i}(z) \, T_{\delta}(k,z) \, j_{\ell}(k,z) \\
\Delta_{\rm g, RSD}^{i}({\ell})  = \int dz \frac{(1+z) n_{\rm g}^{i}(z)}{H(z)} T_{\theta}(k,z)j^{''}_{\ell}(k,z) \\
\Delta_{\rm g, \mu}^{i}({\ell})  = -\ell(\ell+1) \int \frac{dz}{cH(z)} W_{\rm \mu, g}^{i}(z) T_{\phi + \psi}(k,z)j_{\ell}(k,z). 
\end{align}
Here $n_{\rm g}^{i}(z)$ denote the redshift distribution of lens galaxies in the tomographic bin $i$, $j^{''}_{\ell}$ is the second order derivative of the spherical Bessel function, $j_{\ell}$ and $H(z)$ is the Hubble parameter evolution with redshift. Here the magnification efficiency, $W_{\rm \mu, g}^{i}$, can be written as:
\begin{equation}
W_{\rm \mu, g}^{i}(z) = \int_z^{\infty} dz' n_{\rm g}^{i}(z') \frac{\mu_{\rm g}^i(z)}{2} \frac{\chi(z') - \chi(z)}{\chi(z) \chi(z')} 
\end{equation}
The term $b_{\rm g}^{i}(z)$ and $\mu_{\rm g}^i (z)$ encode the evolution of galaxy bias and lens magnification respectively and are the main subjects of investigation of this study. 

The galaxy-galaxy two-point correlations can then be written as:
\begin{align}
C_{\rm gg, obs}^{ij}({\ell}) = \frac{2}{\pi}\int_0^{\infty} \frac{dk}{k}k^3 \, P_{\Phi}(k) \, \Delta^{i}_{\rm g, obs}({\ell}) \, \Delta^{j}_{\rm g, obs}({\ell})
\label{eq:clgg}
\end{align}
where, $P_{\Phi}(k)$ is the primordial power spectrum and is related to the total non-linear matter power spectrum $P_{\delta}$ as $P_{\delta}(k,z_1,z_2) = P_{\Phi}(k) T_{\delta}(k,z_1) T_{\delta}(k,z_2)$. Following the methodology of \cite{Fang2019-jx}, we split the matter power spectra into unequal-time separable linear power spectra that is valid on large scales ($\ell < 200$) and use the fitting formula from halofit in the non-linear scales ($\ell > 200$). We use full non-limber calculations in the linear scales, while resorting to Limber approximation in the non-linear small scales which is good approximation for statistical precision obtainable by LSST \citep{Fang2019-jx}. 

For galaxy lensing and CMB lensing we can write the effective transfer function as following:
\begin{align}
    \Delta_{\rm \gamma_{\rm source}, obs}^{i}({\ell}) = \Delta_{\rm \gamma_{\rm source}, D}^{i}({\ell}) + \Delta_{\rm IA}^{i}({\ell}) \\
    \Delta_{\rm \gamma_{\rm CMB}, obs}^{i}({\ell}) = \Delta_{\rm \gamma_{\rm CMB}, D}({\ell}), \\
\end{align}
where, $\Delta_{\rm \gamma_{\rm source/CMB}, D}^{i}$ is the contribution from the gravitational lensing of source galaxies or CMB, whereas $\Delta_{\rm IA}^{i}$ is an unlensed component and arises due to the intrinsic alignment of the source galaxies in the tidal field of the LSS. We can write the gravitational lensing contribution as \citep{Chisari2019-el}:
\begin{align}
\Delta_{\rm \gamma_{source/CMB}, D}^{i}({\ell}) = \frac{-1}{2} \sqrt{\frac{(\ell+2)!}{(\ell - 2)!}} \int \frac{dz}{c H(z)} W_{\rm {source/CMB}}^{i}(z) {\nonumber} \\ 
 T_{\phi + \psi}(k,z) j_{\ell}(k \chi(z)),
\end{align}
where the lensing efficiency of the source galaxies ($W_{\rm source}^{i}$) and CMB ($W^{\rm CMB}$) can be written as:
and
\begin{align}
W_{\rm source}^{i}(z) = \int_z^{\infty} dz' n_{\rm s}^{i}(z') \frac{\chi(z') - \chi(z)}{\chi(z) \chi(z')} \\
W_{\rm CMB}(z) = \frac{\chi(z_{\star}) - \chi(z)}{\chi(z) \chi(z_{\star})}. 
\end{align}
Here $n_{\rm s}^{i}$ is the redshift distribution of the source galaxies in the tomographic bin $i$ and $z_{\star}$ is the redshift of the surface of last scattering.

We follow the non-linear linear alignment (NLA) model to describe the $\Delta_{\rm IA}^{i}$ component \citep{Hirata2004-qx, Troxel2014-jg}. Although this model fails to describe the intrinsic alignments in small scales \citep{Blazek2017-pj}, it is valid on the scales analyzed in this study \citep{Secco2021-zt}. We can write the contribution from NLA as:
\begin{equation}
\Delta_{\rm IA}^{i}({\ell}) = \sqrt{\frac{(\ell+2)!}{(\ell - 2)!}} \int dz \, n_{\rm s}^{i}(z) \, A_{\rm IA}(z) \, T_{\delta}(k,z) \, \frac{j_{\ell}(k \chi(z))}{|k\chi(z)|^2}.
\end{equation}

The amplitude of the alignment is given by:
\begin{equation}
A_{\rm IA}(z) = \frac{-C_1 \rho_{\rm crit} \Omega_{\rm m}}{G(z)} a_{\rm IA} \bigg( \frac{1+z}{1+z_0} \bigg)^{\eta},
\end{equation}
where we fix $C_1 \rho_{\rm crit} = 0.0134$ as obtained from the SUPERCOSMOS observations \citep{Bridle2007-ws}, $G(z)$ is the matter growth function normalized to unity at $z=0$ and $z_0$ is the pivot redshift that we fix to 0.62 \citep{Troxel2017-cq}. The parameters $a_{\rm IA}$ and $\eta$ are fixed to the median bestfit obtained in the DES-Y3 cosmic shear analysis \citep{Secco2021-zt}.

The cross-correlations between lens galaxy overdensity and lensing of source galaxies or CMB can then be written as:
\begin{align}
C_{\rm g \gamma_{\rm source/CMB}, obs}^{ij}({\ell}) = \frac{2}{\pi}\int_0^{\infty} \frac{dk}{k}k^3 P_{\Phi}(k)\Delta_{\rm g, obs}^{i}({\ell}) \Delta_{\rm \gamma_{\rm source/CMB}, obs}^{j}({\ell}). 
\label{eq:clgammag}
\end{align}
In accordance with \citep{Fang2019-jx} and \citep{Krause2021-fd}, we use Limber approximation when calculating these integrals. 

Our main analysis is performed in the configuration space where the galaxy clustering ($w_{\rm gg}$), galaxy-galaxy lensing ($\gamma_{\rm t, model}$) and galaxy-CMB lensing ($w_{\rm g \kappa_{\rm CMB}}$) are given by:
\begin{align}
w^{ij}_{\rm gg}(\theta) = \sum_{\ell} \frac{2\ell + 1}{4\pi} P_{\ell}({\cos}(\theta)) C_{\rm gg, obs}^{ij}(\ell) \nonumber \\
\gamma^{ij}_{\rm t}(\theta) = \sum_{\ell} \frac{2\ell + 1}{4\pi \ell (\ell + 1)} P^2_{\ell}({\cos}(\theta)) C_{\rm g \gamma_{\rm source}, obs}^{ij}(\ell) \nonumber \\
w^{i}_{g\kappa_{\rm CMB}}(\theta) = \sum_{\ell} \frac{2\ell + 1}{4\pi} P_{\ell}({\cos}(\theta)) C_{\rm g \gamma_{\rm CMB}, obs}^{i}(\ell),\label{eq:wgg_gt_gcmb}
\end{align}
where, $P_{\ell}$ and $P^2_{\ell}$ are the Legendre and the associated Legendre polynomials. 

The galaxy-galaxy lensing observable is a non-local statistic. This results in $\gamma^{ij}_{\rm t}(\theta)$ being sensitive to scales $\theta' <= \theta$, where theoretical modeling uncertainties are high. We remove the sensitivity to these highly non-linear small scales by removing the relevant mode \citep{Baldauf2009-gy}. We follow the scheme of \cite{MacCrann2019-wp}, where the small scales are shown to contribute as an effective point mass. Therefore, the galaxy-galaxy signal transforms to:
\begin{equation}
\gamma^{ij}_{\rm t, wPM}(\theta) = \gamma^{ij}_{\rm t}(\theta) + \frac{M^{ij}}{\theta^2},
\end{equation}

\begin{equation}
M^{ij} = \int dz_{\rm g} \, dz_{\rm s} \, n^i_{g} \, n^i_{s} \, \frac{B^{ij}}{\Sigma^2_{\rm crit}(z_{\rm g} ,z_{\rm s}) D^2_{\rm A}(z_{\rm g})} 
\end{equation}

We use completely uninformative prior on the parameters $B^{ij}$ which capture the strength of non-linearity in the small scales. As described in the following sub-section, we marginalize over these parameters analytically by modifying the covariance. Note that this scheme also accounts for the evolution of the point mass parameters with redshift. 

\subsection{Covariance}

For covariance calculations, we use the Gaussian approximation which is sufficient for current generation surveys \citep{Friedrich2020-rh}. For this study, we assume this approximation to remain valid for next generation surveys as well and leave a detailed analysis for future study. 

We define the different real-space correlation functions as
\begin{equation}
\xi^{ij}_{AB} = \sum_{\ell} \frac{2\ell + 1}{4\pi} F^{AB}_{\ell}(\theta) C_{AB}^{ij}(\theta),
\end{equation}
where, $AB \in {gg, g \gamma_{\rm source}, g \gamma_{\rm CMB}}$ and the $F^{AB}_{\ell}(\theta)$ can be read off from Eq.~\ref{eq:wgg_gt_gcmb}. 
Note that we additionally include exact bin averaging when estimating the $F^{AB}_{\ell}(\theta)$, as described in \citep{Friedrich2020-rh}. 
The covariance then becomes:
\begin{align}
\mathbfcal{C} = {\rm Cov}[\xi^{ij}_{AB}(\theta_1), \xi^{kl}_{CD}(\theta_2)] = \sum_{\ell_1 \ell_2} \frac{(2\ell_1 + 1)(2\ell_2 + 1)}{(4\pi)^2} {\nonumber}\\
F^{AB}_{\ell_1}(\theta_1) F^{CD}_{\ell_2}(\theta_2) {\rm Cov}[\hat{C}^{ij}_{AB}(\ell_1), \hat{C}^{kl}_{CD}(\ell_2)]
\end{align}

Here, ${\rm Cov}[\hat{C}^{ij}_{AB}(\ell_1), \hat{C}^{kl}_{CD}]$ is the covariance between probes in the Fourier space. Note that here, $\hat{C}^{ij}_{AB}(\ell_1) = {C}^{ij}_{AB}(\ell_1) + \mathcal{N}^{ij}_{AB}(\ell_1)$ includes the noise contribution $\mathcal{N}^{ij}_{AB}$. The noise contribution in case of:
\begin{itemize}
	\item galaxy-galaxy correlation is $\mathcal{N}^{ij}_{\rm gg}(\ell_1) = \frac{\delta_{ij}}{\bar{n}^i_{\rm g}}$
	\item for galaxy lensing-galaxy lensing correlation is $\mathcal{N}^{ij}_{\rm \gamma_{\rm source}\gamma_{\rm source}}(\ell_1) = \frac{\delta_{ij} \sigma^2_{{\rm e},i}}{\bar{n}^i_{\rm s, eff}}$
	\item for CMB lensing-CMB lensing correlation is $\mathcal{N}^{ij}_{\rm \gamma_{\rm CMB}\gamma_{\rm CMB}}(\ell_1) =  \mathcal{N}_{\rm CMB}(\ell_1)$. We use the noise curve for combined SPT+Planck as shown in the Figure.~3 of \cite{Omori2022-ve}. 
\end{itemize}

Note that the noise contribution vanishes in case $A \neq B$. For both LSST-Y1 and LSST-Y10, we show the inputs in Table~\ref{tab:cov_settings} \citep{The_LSST_Dark_Energy_Science_Collaboration2018-ky}.

As mentioned in the previous section,  galaxy-galaxy lensing is a non-local statistic. Therefore, a point-mass marginalization scheme is important to ensure theoretical model uncertainties are under control. We include this effect by following the procedure outlined in \cite{MacCrann2019-wp} and modifying the covariance matrix which is equivalent to an analytical marginalization. We assume that the point-mass contribution can also evolve with redshift hence each lens-source redshift combination will be sensitive to a different value of point mass. Therefore, as described in \cite{MacCrann2019-wp}, the effective change in the covariance matrix is given by:   
\begin{equation}
    {\mathbfcal{C}}^{-1}_{\rm wPM} = {\mathbfcal{C}}^{-1} - {\mathbfcal{C}}^{-1} {\mathbfcal{U}} ({\mathbfcal{I}} + {\mathbfcal{U}}^{\rm T} {\mathbfcal{C}}^{-1} {\mathbfcal{U}})^{-1} {\mathbfcal{U}}^{\rm T} {\mathbfcal{C}}^{-1}
\end{equation}

Here ${\mathbfcal{C}}^{-1}$ is the inverse of the halo-model covariance as described above, $\mathbfcal{I}$ is the unity matrix and $\mathbfcal{U}$ is a $N_{\rm data} \times N_{\rm p}$ matrix where $N_{\rm p} = n_{\rm lens}\times n_{\rm source}$ with $n_{\rm lens}$ and $n_{\rm source}$ being the number of lens and source redshift bins respectively. The $p$-th column of this $\mathbfcal{U}$ matrix is given by $\sigma_{B} \times \vec{t}^{ij}$. Here $\sigma_{B}$ is the standard deviation of the Gaussian prior on point mass parameter $B$ and we put a large prior of $\sigma_{B} = 10^5$ (which translates to the effective mass residual prior of $10^{17}M_{\odot}/h$.) \citep{Pandey2021-jc}. The matrix $\vec{t}^{ij}$ is given by:

\begin{equation}
    \bigg(\vec{t}^{ij} \bigg)_{a} = \begin{cases}
\theta_{a}^{-2} & \parbox{5cm}{if $a$-th element correspond to $\gamma_{\rm t}$ and if lens-source redshift pair for $a$-th element $= ij$} \\  
\\
0 &\text{otherwise}
\end{cases}
\end{equation}

\begin{figure}
	\includegraphics[width=0.95\columnwidth]{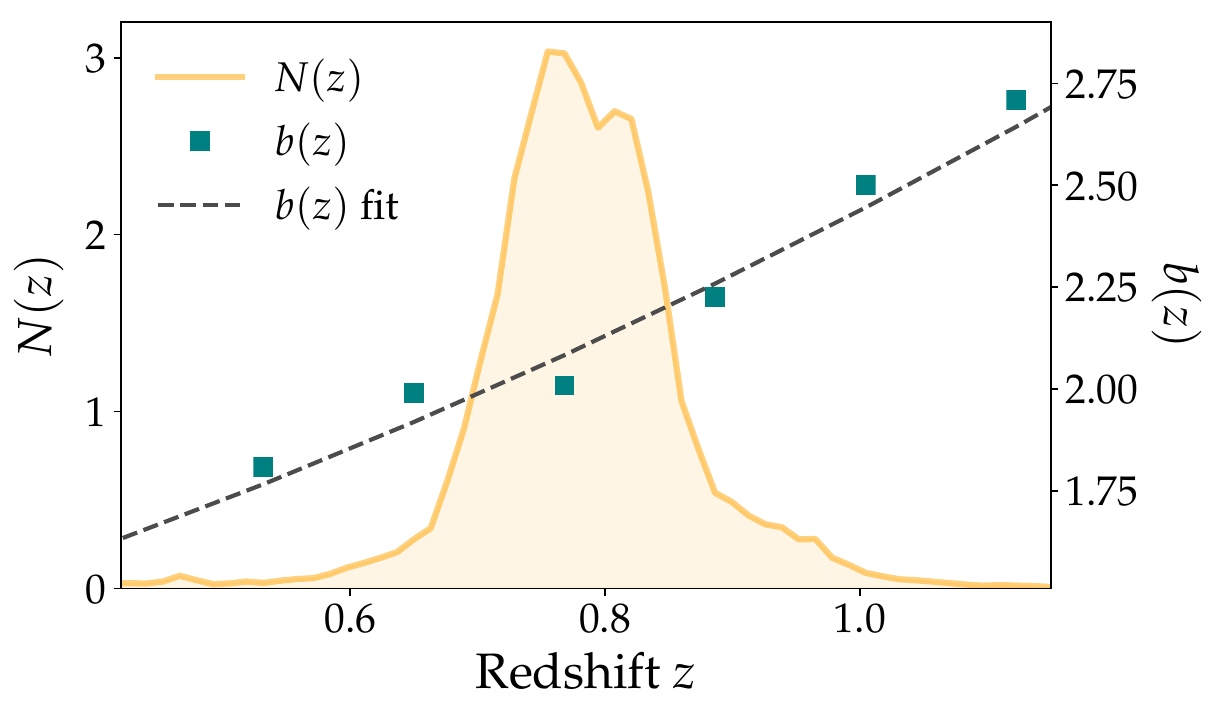}
	\caption[]{Comparison of large scale bias and analytic powerlaw distribution of MagLim galaxies in the \mice simulation for a given lens redshift bin. The powerlaw distribution follows Eq.~\ref{eq:bz} with $\alpha=2.0$ and $b_0=2.1$. 
	}
	\label{fig:bz_sims}
\end{figure}

\begin{figure}
\includegraphics[width=\columnwidth]{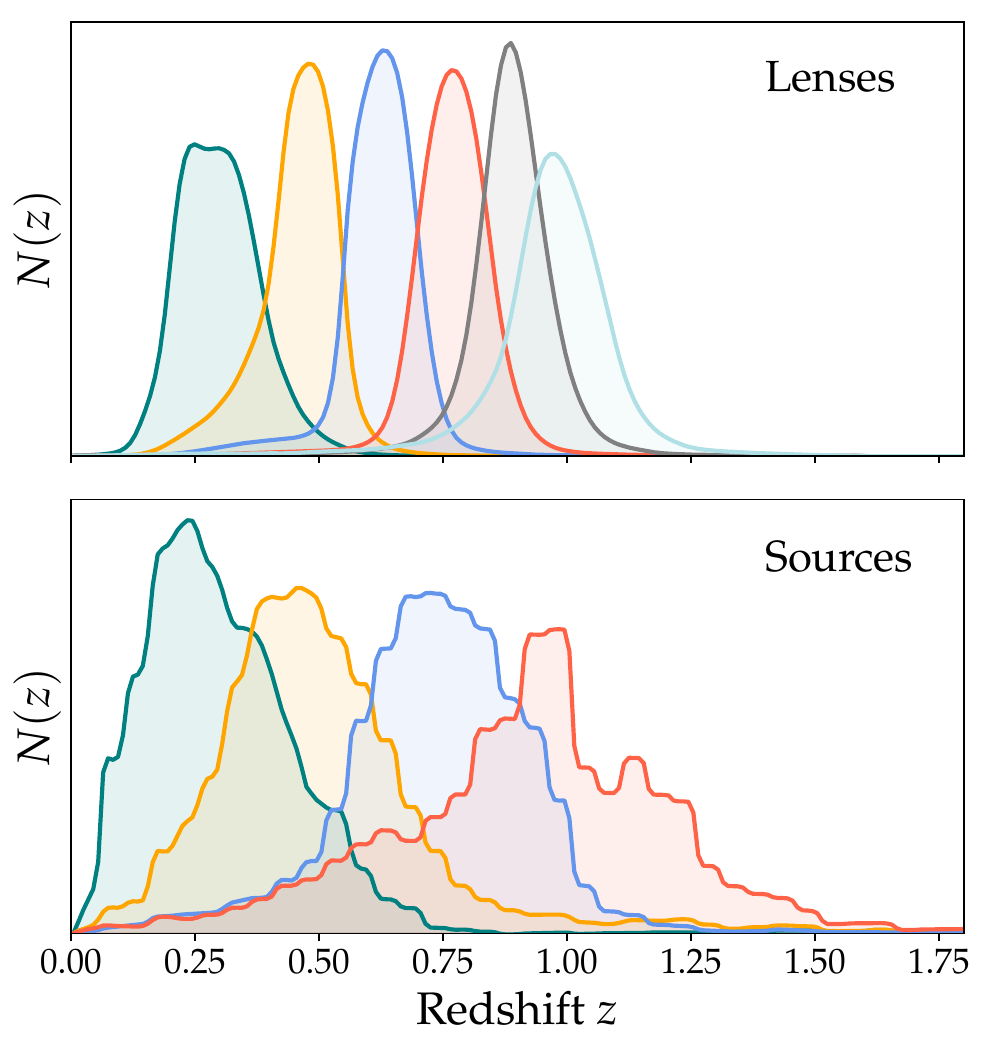}
\caption[]{Redshift distributions of the lens and source samples used in this work, based on those obtained in the DES Year 3 fiducial analysis \citep{DES_Collaboration2021-gz}.
}
\label{fig:nz}
\end{figure}

\section{Simulated tests}
\label{sec:sim_settings}

\begin{figure*}
\includegraphics[width=\textwidth]{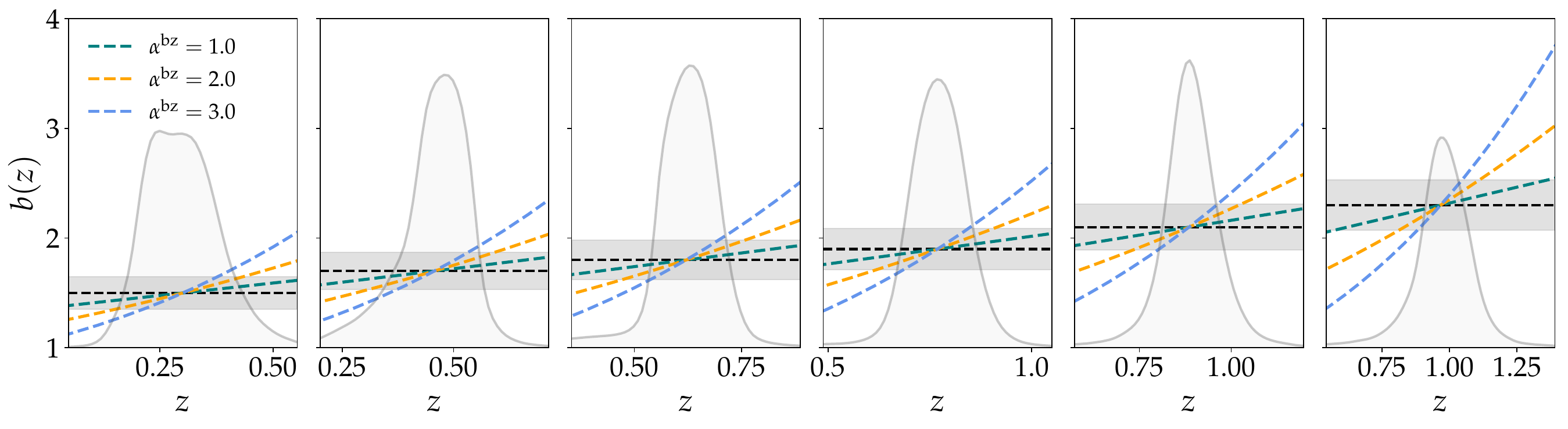}
\caption[]{Comparing the evolution of large scale bias for the tomographic bins shown in Fig.\ref{fig:nz}. For each panel, the central value corresponds to the mean redshift of the bin and the redshift range spans approximately $3\sigma$ of the variance in $n(z)$. 
}
\label{fig:bz_powerlaw}
\end{figure*}

We aim to test the impact of evolving astrophysical effects, like galaxy bias and lens magnification, on the measurements of galaxy clustering, galaxy-galaxy lensing and galaxy-CMB lensing. Our analysis is aimed primarily at Stage IV surveys, Rubin-LSST and Euclid in particular. While the expected redshift distributions for LSST (which is also expected to influence those of Euclid via its shared data in optical bands) have been published, we expect that the distributions estimated from data will differ. In particular, recent developments in minimizing biases in the mean and widths of estimated redshift distributions rely on techniques such as self-organizing maps (SOMs) in color space \citep{Carrasco_Kind2013-hd, Sanchez2018, Sanchez2020, Buchs2019-zo}. The current release of DES-Y3 constitutes a data-based example of such estimation, providing realistic tails in the derived redshift distributions \citep{Myles2020-pb}. We model the evolution for imaging galaxy samples via the \mice simulations \citep{Crocce2013-vx} catalog of DES-Y3 like MagLim galaxies. These galaxies are selected by imposing a redshift dependent cut on their simulated $i$-band magnitudes \citep{Porredon2020-aa}. This effectively leads to a complicated, redshift-dependent color selection. The photometric redshifts are estimated using the DNF algorithm \citep{De_Vicente2015-wb}, as used in the data analysis as well. We show the redshift distribution of the galaxies selected using $0.65 < z_{\rm DNF} < 0.85$. We identify the host halos of each galaxy in this bin and using their true redshift and halo mass, we estimate the average galaxy bias using the Eq.~\ref{eq:bz_HM}. We fit this redshift evolution of bias using a power-law phenomenological function as follows:
\begin{equation}\label{eq:bz}
    b^{i}_{\rm g}(z) = b^{i}_{\rm const} \frac{1 + \frac{1}{3}((1 + z)^{\alpha^{i}_{\rm bz}} - 1)}{1 + \frac{1}{3}((1 + \bar{z})^{\alpha^{i}_{\rm bz}} - 1)},
\end{equation}
where $\bar{z}$ is the mean redshift of the tomographic bin $i$. Figure \ref{fig:bz_sims} shows how this function provides a good fit to the evolution of galaxy bias in the simulation.

We model the evolution of the magnification coefficients using similar functional form:
\begin{equation}
    \mu^{i}_{\rm g}(z) = \mu^{i}_{\rm const} \frac{1 + \frac{1}{3}((1 + z)^{\alpha^{i}_{\rm \mu z}} - 1)}{1 + \frac{1}{3}((1 + \bar{z})^{\alpha^{i}_{\rm \mu z}} - 1)},
\label{eq:muz}    
\end{equation}

We see that this functional form provides a good fit to the measured bias values \citep{Elvin-Poole2022-vg}. In our analysis, $b^{i}_{\rm const}$, $\mu^{i}_{\rm const}$, $\alpha^{i}_{\rm bz}$ and $\alpha^{i}_{\rm \mu z}$ are the parameters of interest in this study.

For our simulated tests, we use the same redshift distributions as obtained for the MagLim sample in DES-Y3 analysis \citep{DES_Collaboration2021-gz}. The redshift distribution for lens and source galaxies is shown in Fig.~\ref{fig:nz}. For these redshift distributions, we show different powerlaw predictions for the galaxy bias evolution in Fig.~\ref{fig:bz_powerlaw}. Note that Rubin LSST redshift distributions will differ from the ones we use, but we prioritized the use of real-data redshift distributions for this work. 

We estimate the covariance for LSST-Y1 and LSST-Y10 like setup \citep{The_LSST_Dark_Energy_Science_Collaboration2018-ky}. For this, we fix $f_{\rm sky}=0.4$. We vary the shot-noise and shape-noise for the two cases, by changing the observed number density of the lens and source galaxies, $n^i_{\rm g}$ and $n^i_{\rm eff}$ respectively as shown in Table~\ref{tab:cov_settings}.

We also show results for the case when same sample is used for both lens and source galaxies. In this case, we fix the lens galaxy distribution to be same as the source galaxies as shown in the bottom panel of Fig.~\ref{fig:nz}. Moreover, we fix the $\bar{n}^i_{\rm g}$ to be same as $\bar{n}^i_{\rm eff}$ in Table~\ref{tab:cov_settings} when estimating the covariances for this analysis setup. 

Note that in this analysis we make use of the \texttt{CosmoSIS}\footnote{https://github.com/joezuntz/cosmosis/} likelihood and sampling framework \citep{Zuntz_2015}. For sampling, we use the \texttt{Multinest} nested sampling procedure as described in \cite{Feroz:2009:MNRAS:}.

\begin{table}
\begin{tabular}{ |p{3cm}||p{2cm}|p{2cm}| }
 \hline
 Noise estimates & $\bar{n}^{i \in [1,..,6]}_g$ & $\bar{n}^{j \in [1,..,4]}_{\rm eff}$\\
 \hline
 Y1   & 0.3    & 3.5 \\
 Y10&   3.0  & 7.0  \\
 \hline
\end{tabular}
\caption{{We describe the effective galaxy densities for lens and source galaxy samples used in the covariance estimates for the Rubin-LSST Y1 and Y10 configurations explored in this analysis.}}
\label{tab:cov_settings}
\end{table}

\section{Results}
\label{sec:results}
\subsection{Impact on the measurements}


\begin{figure*}
\centering
\includegraphics[width=\textwidth]{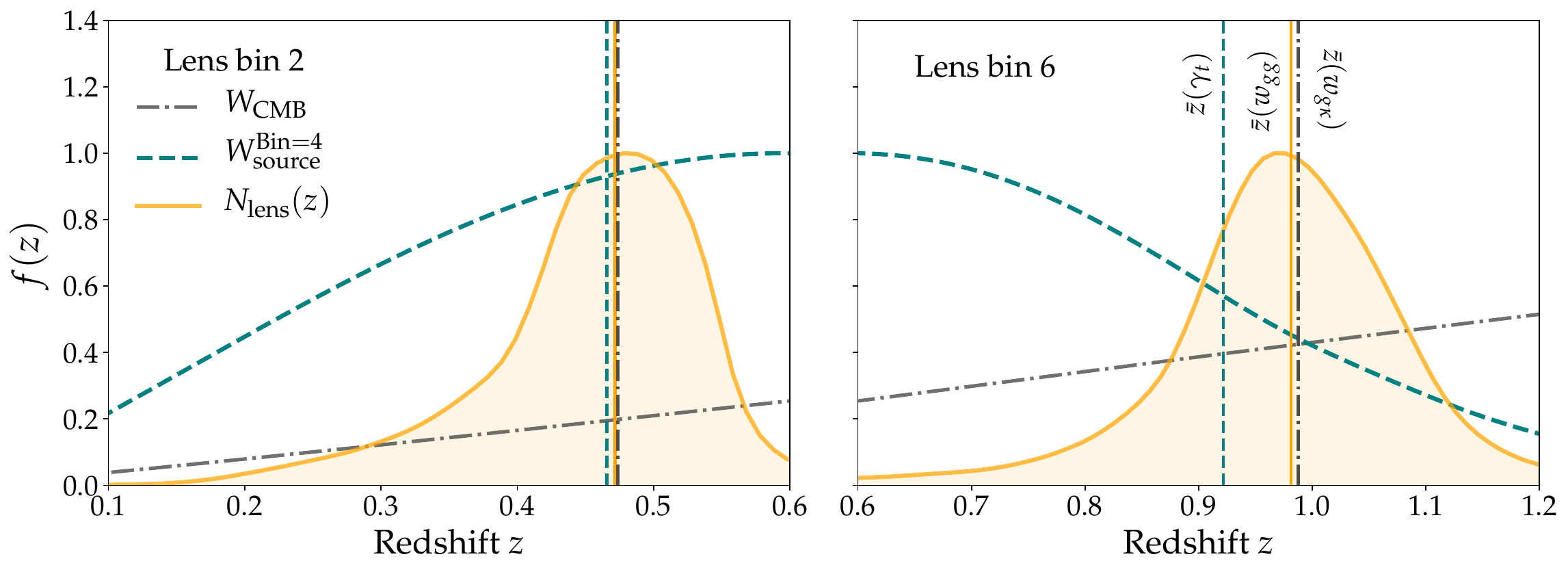}
\caption[]{
In this figure we show the lensing efficiency for CMB photons ($W_{\rm CMB}$), for photons emitted by source galaxies corresponding to fourth redshift bin ($W_{\rm source}^{\rm Bin=4}$) and redshift distribution of lenses ($n_{\rm lens}$) corresponding to the second (sixth) tomographic on the left (right) panel. With vertical lines, we show the mean redshift that contributes to the auto-correlations of lens galaxies (blue line), galaxy-galaxy lensing (black dashed line) and galaxy-CMB lensing (red dot-dashed line). We see that while for the second lens tomographic bin, all three probes receive contribution from similar mean redshift, for the sixth tomographic bin, galaxy-galaxy lensing receives contribution from significantly lower redshift than the other two probes. 
}
\label{fig:zmean_probes}
\end{figure*}

\begin{figure*}
\includegraphics[width=\textwidth]{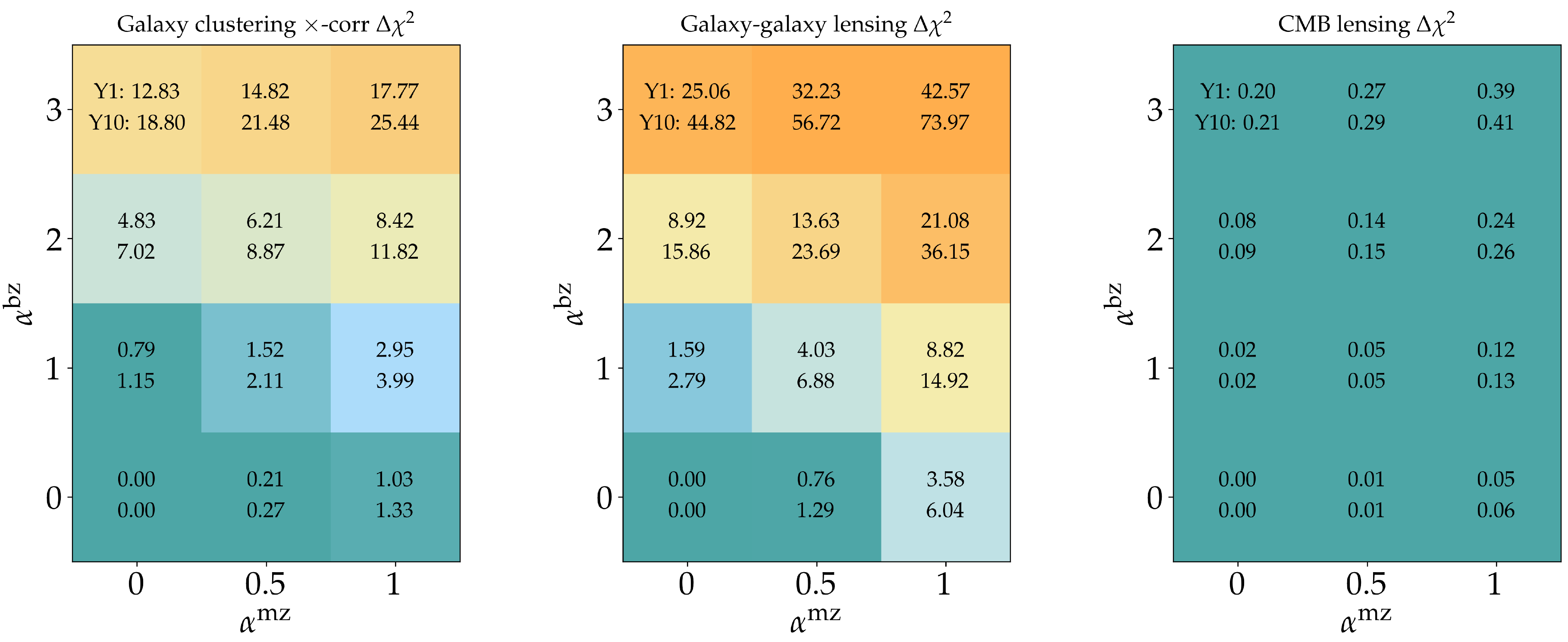}
\caption[]{In this figure, we show $\Delta \chi^2$ between the fiducial simulated datavector assuming no evolution of galaxy bias and lens magnification and a simulated datavector with evolving bias and lens magnification with powerlaw coefficient $\alpha^{\rm bz}$ and $\alpha^{\rm mz}$ respectively. On left panel we show the forecasted $\Delta \chi^2$ in the cross-tomographic bin galaxy clustering correlations, the central panel corresponds to galaxy-galaxy lensing, and the right panel to the galaxy-CMB lensing correlations. The top (bottom) numbers in each block corresponds to assuming LSST-Y1 (Y10) covariance.
}
\label{fig:chi2_values}
\end{figure*}

We first estimate the impact of the evolution of both galaxy bias and lens magnification on the measurements of galaxy clustering, galaxy-galaxy lensing and galaxy-CMB lensing. In Fig.~\ref{fig:zmean_probes}, we show the origin of impact on these different measurements. As we can see from Eq.~\ref{eq:clgg} and Eq.~\ref{eq:clgammag}, while the galaxy-galaxy auto-correlations are sensitive to $(n^{i}_{\rm g})^2$, the cross-correlations of galaxies with galaxy lensing or CMB lensing will be sensitive to $n^i_{\rm g} W^j_{\rm source/CMB}$. The effective redshift of these probes will hence be different, depending on the shape of the kernels $W^j_{\rm source/CMB}$. In Fig.~\ref{fig:zmean_probes}, we show this effective redshift that each probe is sensitive to using the vertical lines. We can clearly see that while the galaxy clustering and galaxy-CMB lensing correlations probe similar mean redshifts, the galaxy-galaxy lensing correlations are sensitive to lower redshift, especially for the case of lens bins at higher redshift. This discrepancy is due to the shape of the galaxy lensing kernel, which varies rapidly close to the redshift of the source bin. In the case of an evolving galaxy bias, this difference in the effective mean redshift would lead to difference in the amplitude of the signal. We can similarly argue that as the galaxy-galaxy cross-bin correlations between bins $i$ and $j$ would be sensitive to $(n^{i}_{\rm g} n^{j}_{\rm g})$, they would similarly probe different effective redshift compared to the auto-correlations. 

\begin{figure}
\includegraphics[width=0.95\columnwidth]{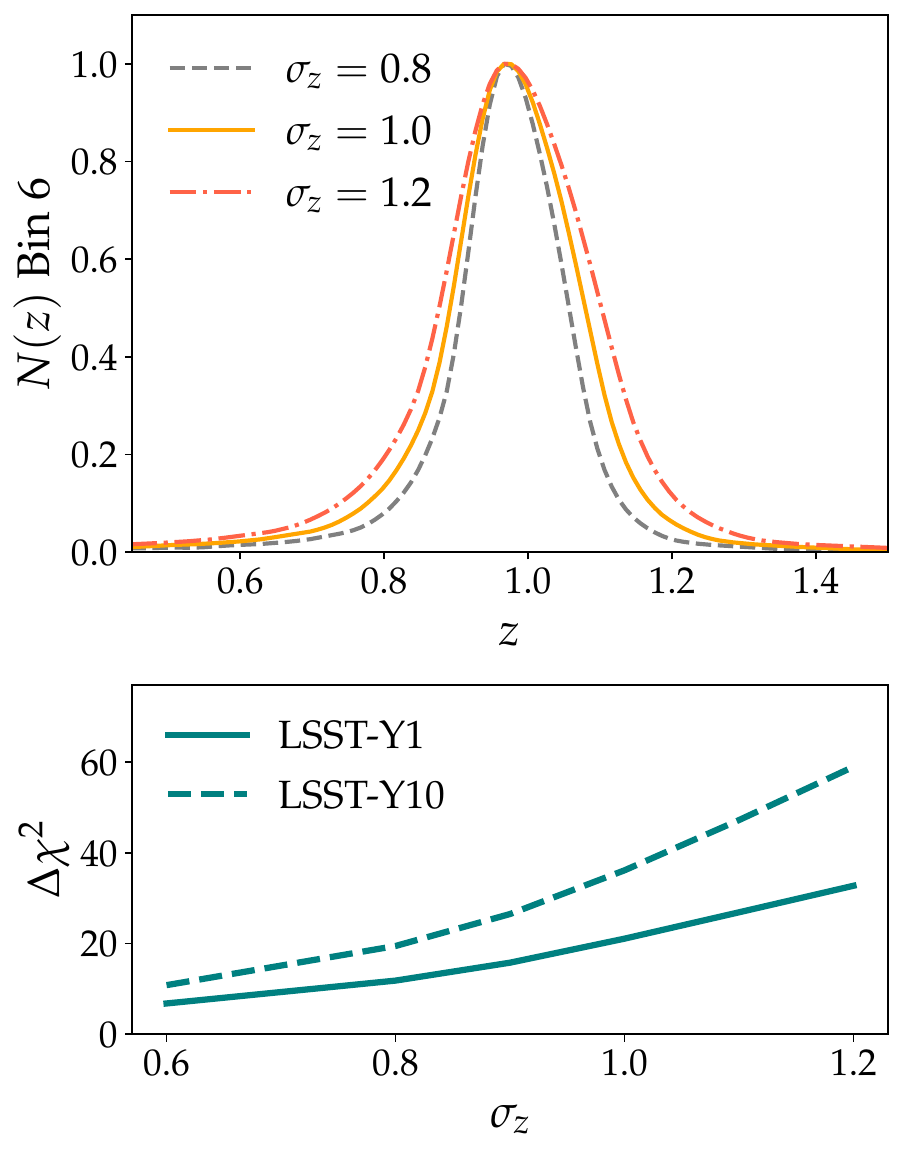}
\caption[]{In this figure we test the impact of the width of redshift distribution of the lens galaxies. In the top panel we show how the $n(z)$ of the galaxies in the sixth tomographic bin changes with changing the width parameter, $\sigma_z$. In the bottom panel we show the impact this has on the $\Delta \chi^2$ between galaxy-galaxy lensing datavector assuming a constant bias and magnification, and one with evolution coefficient $\alpha^{\rm bz} = 2.0, \alpha^{\rm mz} = 1.0$ respectively. The solid (dashed) line corresponds to LSST- Y1 (Y10) covariance.
}
\label{fig:sigz}
\end{figure}

In Fig.~\ref{fig:chi2_values} we show the impact of galaxy bias and lens magnification evolution on the different measurements we consider, quantified by the $\Delta \chi^2$ between the measurements with and without these effects. Note that these $\Delta \chi^2$ use the same scale cuts as described in DES 5$\times$2 analyses \citep{Abbott2022-xa, Omori2022-ve}. These roughly correspond to the minimum angular scales corresponding to $8\  {\rm Mpc}/h$ and $6\  {\rm Mpc}/h$ physical separation for galaxy clustering and galaxy-galaxy lensing respectively. We refer the reader to \cite{Omori2022-ve} for details regarding obtaining the CMB lensing scale cuts. The average values for galaxy bias and magnification, $b_0$ and $\mu_0$, are chosen such that the galaxy clustering auto-correlations match. We make this choice because we explicitly aim to test the consistency of galaxy auto-correlations and cross-correlations under the same modeling assumption. Moreover, the detection significance of the galaxy-galaxy auto-correlations is significantly higher than that of the other probes considered here. Therefore, they typically fix the average galaxy bias $b_0$. Furthermore, we typically have access to tight constraints on the mean lens magnification values from empirical methods using artificial galaxy injections, like Balrog \citep{Everett2020-ru} in DES, that are broadly used in imaging survey analyses \citep{Elvin-Poole2022-vg}. 

As shown in Fig.~\ref{fig:chi2_values}, both the galaxy-galaxy lensing and galaxy clustering cross tomographic bin correlations are heavily impacted by the astrophysical evolution. In \S~\ref{app:constrain_bzmz}, we constrain this evolution by doing a joint analysis of all the probes, even when marginalizing over various cosmological and systematics parameters with wide prior.

\subsubsection{Sensitivity to lens redshift distributions}
The effects described in this work are caused by the evolution of astrophysics within the finite width of lens redshift distributions. Therefore, we expect such effects to have a smaller impact in the case of narrower lens redshift distributions. We change the width, we vary the $\sigma_z$ parameter for each tomographic bin, which changes the redshift distribution as \citep{Cawthon2020-hm}:
\begin{equation}
    n^i(z) \longrightarrow \frac{1}{\sigma_z} n\bigg( \frac{z - z^i_{\rm mean}}{\sigma_z} + z^i_{\rm mean} \bigg),
\end{equation}
where, $z^i_{\rm mean}$ is the mean redshift of the tomographic bin $i$. In the top panel of Fig.~\ref{fig:sigz}, we show the impact of changing $\sigma_z$ on the redshift distribution of the sixth tomographic bin. 

In the bottom panel of the Fig.~\ref{fig:sigz} we show the impact of changing $\sigma_z$ on the $\Delta \chi^2$  for galaxy-galaxy lensing as described in the previous section. We fix the powerlaw coefficients of bias and magnification evolution to $\alpha^{\rm bz} = 2.0$ and $\alpha^{\rm mz} = 1.0$ respectively. As shown in Fig.~\ref{fig:bz_sims}, $\alpha^{\rm bz} = 2.0$ provides a good fit to the DES-like galaxies. As we lack all the necessary data products to make a similar plot for the magnification coefficients, we approximate its evolution using the bestfit coefficients obtained from the DES analysis. \cite{Elvin-Poole2022-vg} provide bestfit coefficients for multiple tomographic bins. We find that $\alpha^{\rm mz} = 1.0$ provides a conservative fit to those constraints. We find that shrinking the lens redshift distributions, at fixed mean redshift, by approximately a factor of two would bring down the total $\Delta \chi^2$ by more than a factor of three. 

This provides motivation for defining galaxy samples with precise photometric redshift estimates, for example using the \redmagic algorithm \citep{Rykoff2016-iy}. These algorithms typically rely on finding red galaxies which have tight relation with redshift. However, we note that as these samples use explicit color-cuts, they are usually more susceptible to photometric measurement issues due to observing conditions \citep{Pandey2021-jc}.

\begin{figure*}
\includegraphics[width=\textwidth]{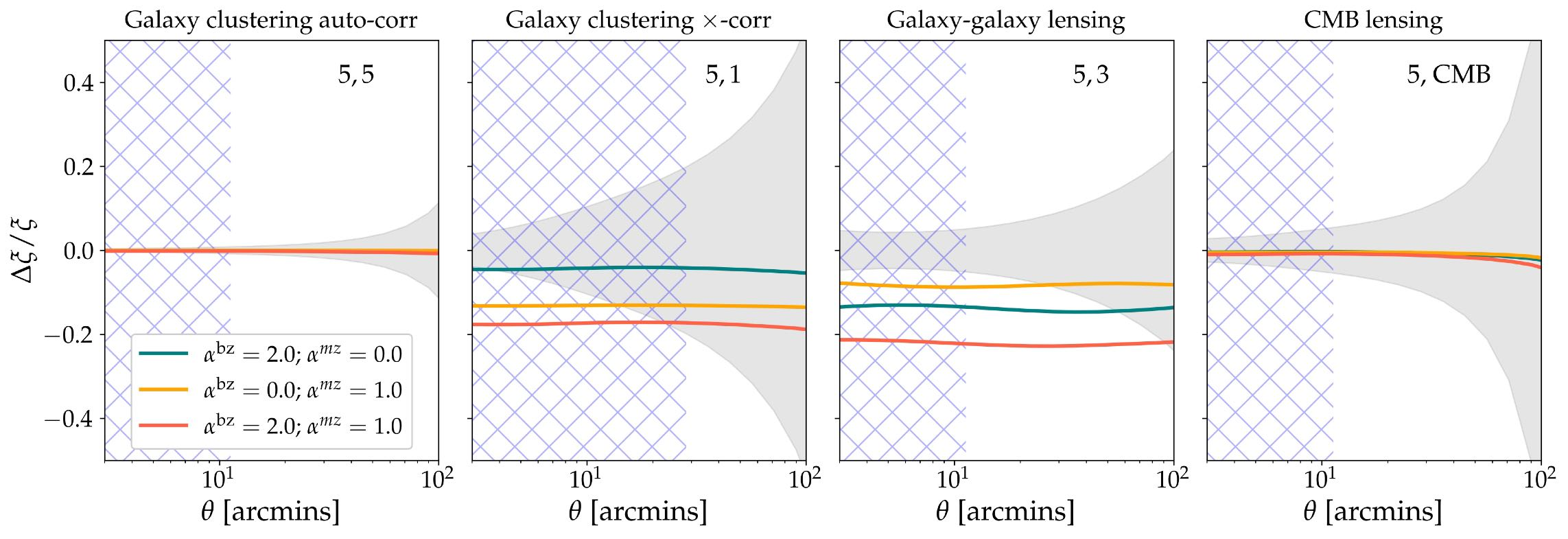}
\caption[]{
Comparing the residuals of the datavector. The hatched area shows the scales excluded from the $\Delta \chi^2$ calculations in Fig.~\ref{fig:chi2_values}. 
}
\label{fig:resid_DV}
\end{figure*}

\begin{figure*}
\includegraphics[width=\textwidth]{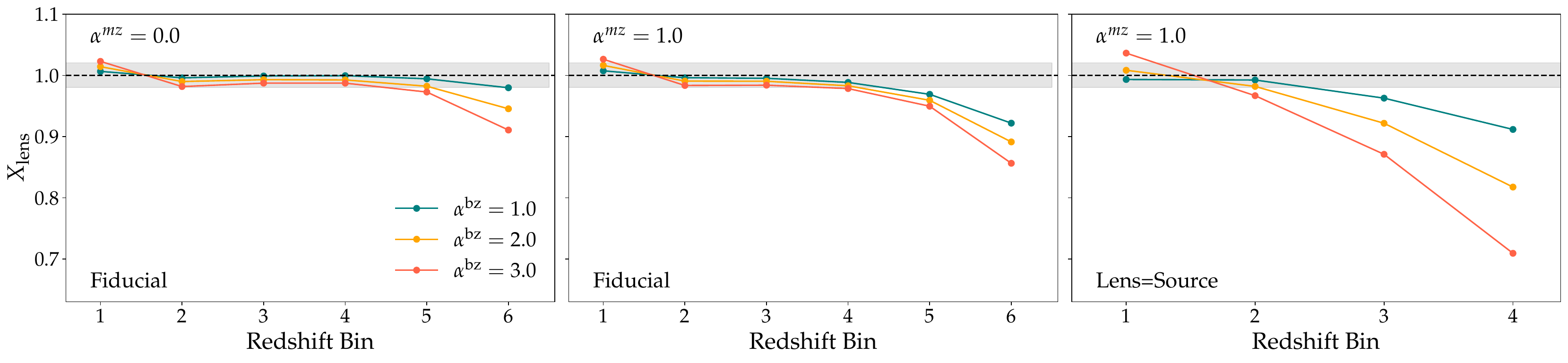}
\caption[]{
Comparing how the inferred $X_{\rm lens}$ evolves with redshift bin for different bias evolution profiles. On the left panel we fix the lens magnification evolution to zero while on the central panel we fix the power-law index of lens magnification evolution to 1.0. In the right panel, we show the case of using same galaxy sample for both the lens and source galaxies. The gray band corresponds to 2\% departure from the fiducial value of $X_{\rm lens}=1$, which approximately encapsulates the small scale non-linear effects at our scale cuts.
}
\label{fig:Xinfer}
\end{figure*}

\subsection{Apparent cosmological de-correlation}
In order to further understand the impact of astrophysical evolution considered here, we look at the residuals of the predicted measurements. In the Fig.~\ref{fig:resid_DV}, we find that remarkably, the residuals behave like a constant multiplicative factor, with little angular scale dependence. Therefore, we parameterize these as $\gamma_{\rm t, scaled} = X_{\rm lens} \gamma_{\rm t, unscaled}$. In Fig.~\ref{fig:Xinfer} we show the mean values of the recovered $X_{\rm lens}$ parameters. Note that in order to infer these, we jointly fit the galaxy-galaxy auto-correlations and galaxy-galaxy lensing datavector with a constant bias model, free $X^i_{\rm lens}$ and $b^i_{\rm const}$ parameters. We do not show the errorbars since those will depend upon the exact analysis settings such as free systematics parameters and prior on those. However, as these are noiseless datavectors, the mean values shown in the plot are robust. We find a de-correlation up to 15\% depending upon the assumed evolution coefficient of bias and magnification values. In Fig.~\ref{fig:Xinfer}, we also show the expected impact of small scale non-linearity with a gray band. For example, as found in previous simulated analysis with photometric galaxy samples \citep{Pandey2020-xu, Goldstein2021-lh}, we expect approximately 2\% effect of non-linear galaxy biasing. These analyses use simulated mock galaxy catalogs relevant to DES and Rubin surveys and show that the cross-correlation coefficient between galaxy matter to deviate from expected value of 1 by less than 2\%. This is significantly smaller than the 15\% effect of evolution, especially at higher redshift.

\subsection{Lens=Source distribution}
In this subsection, we make the lens sample the same as the source galaxy sample. This configuration significantly improves the number density of lens galaxies and the associated redshift characterization \citep{2022MNRAS.509.5721F, 2020JCAP...12..001S}, at the expense of broader redshift distributions 
However, given broader redshift distributions, this configuration is more prone to effects caused by the evolution of galaxy biases or magnification coefficients. We find that assuming $\alpha^{\rm bz} = 2.0$ and $\alpha^{\rm mz}=1.0$ results in a $\Delta \chi^2 \approx 200$ for the galaxy-galaxy lensing only measurement in this case. The inferred de-correlation amplitude for this case is shown in the right panel of Fig.~\ref{fig:Xinfer}. The de-correlation reaches approximately 30\% in this case. While both Euclid and LSST surveys have given serious consideration to using same redshift distribution for lens and source galaxies \citep{Euclid_Collaboration2019-aw, The_LSST_Dark_Energy_Science_Collaboration2018-ky}, our results show that it may be worth revisiting this choice in favor of narrower lens redshift bins.

\begin{figure*}
\begin{minipage}{0.33\textwidth}
    \centering
    \includegraphics[width=\linewidth]{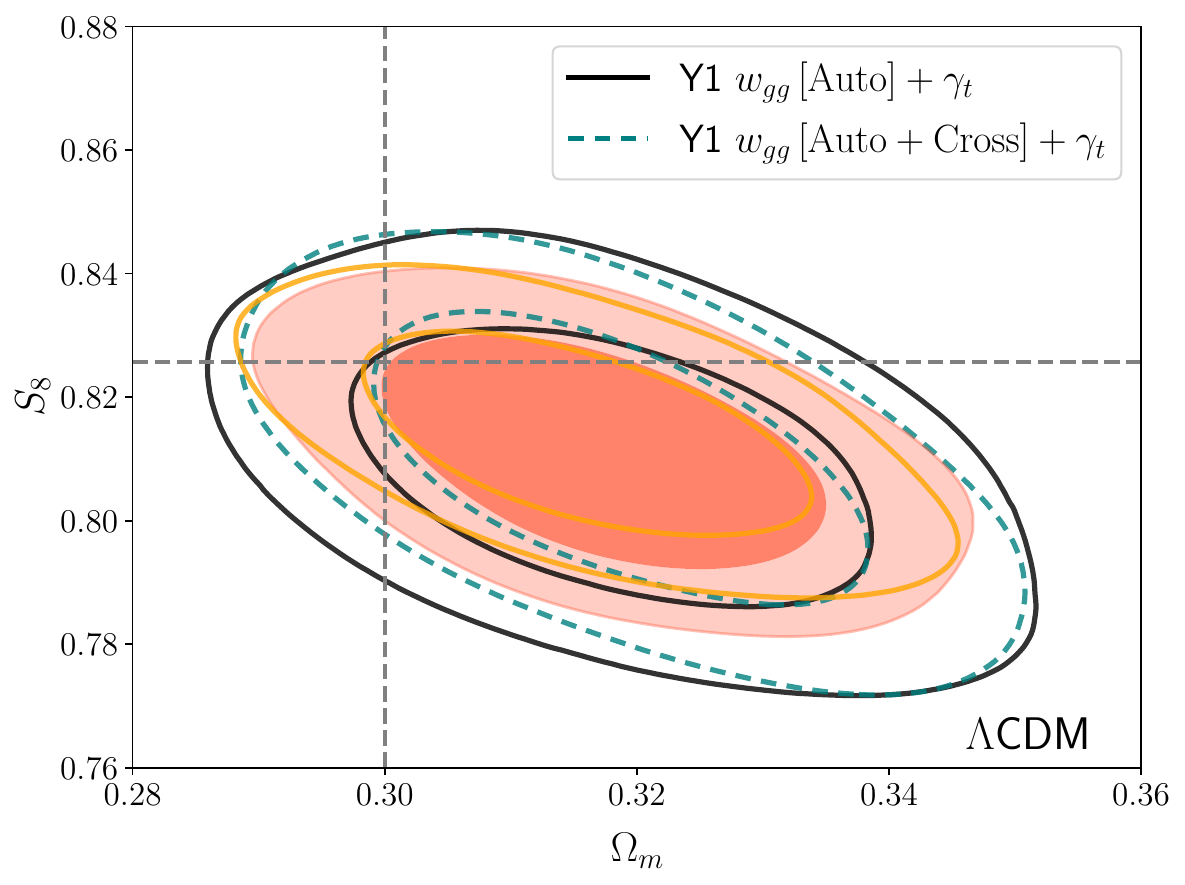}
\end{minipage}%
\begin{minipage}{0.33\textwidth}
    \centering
    \includegraphics[width=\linewidth]{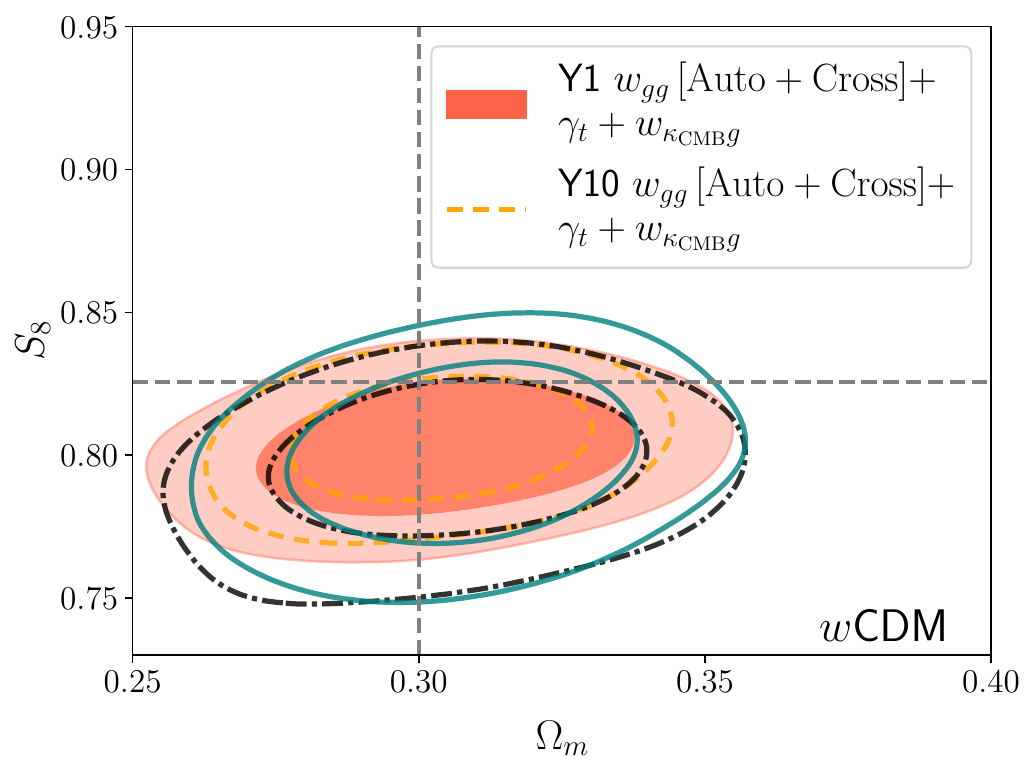}
\end{minipage}%
\begin{minipage}{0.33\textwidth}
    \centering
    \includegraphics[width=\linewidth]{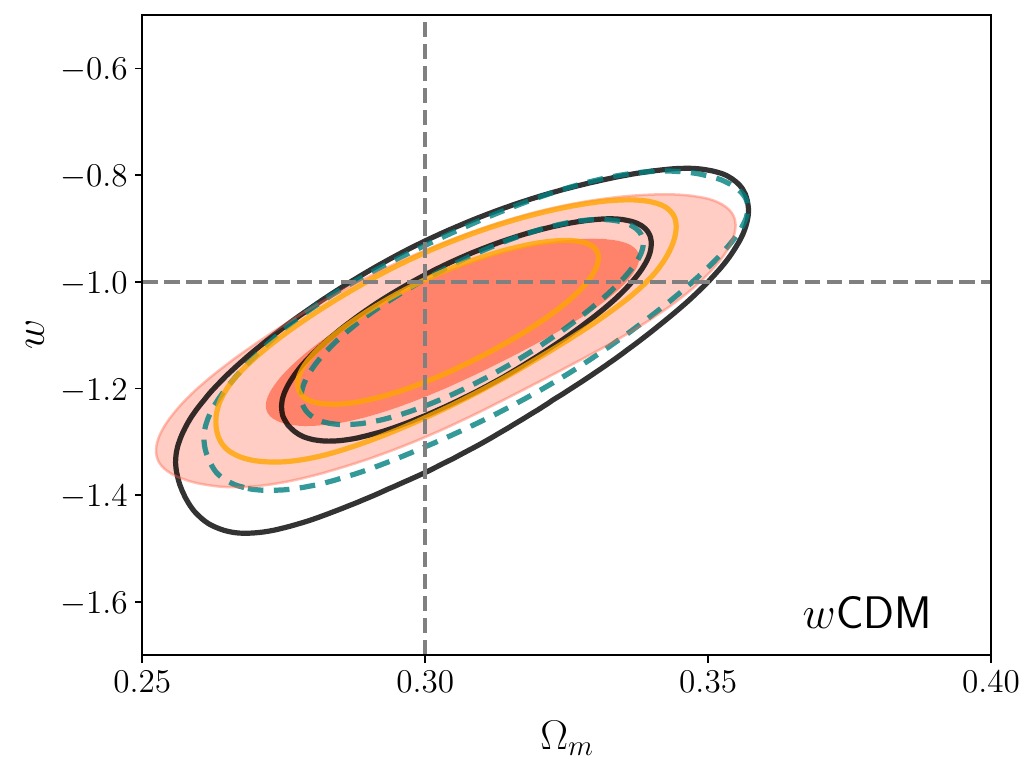}
\end{minipage}
\caption[]{
    Simulated forecast analyzing a datavector (with probes mentioned in the legend) constructed using evolving bias and magnification with a constant parameter model. In addition to the parameters of analysis presented in \cite{DES_Collaboration2022-yt}, this analysis varies lens magnification parameters as well. We present both the $\Lambda$CDM analysis (left panel) and $w$CDM analysis (middle and right panel). 
}
\label{fig:cosmo_const}
\end{figure*}

\subsection{Cosmological impact of the evolution and constraining the parameters}
\label{app:constrain_bzmz}

We estimate the impact of the evolution of galaxy bias and lens magnification on the cosmological parameters in Fig.~\ref{fig:cosmo_const}. In these results we assume the survey specficiations mentioned in \S~\ref{sec:sim_settings}. We analyze a simulated datavector obtained using $\alpha^{\rm bz} = 2.0$ and $\alpha^{\rm mz}=1.0$ with a constant galaxy and lens magnification model. We analyze both constraints using both the Y1 and Y10 simulated covariances. We marginalize over all the cosmological and astrophysical systematics parameters as described in \cite{DES_Collaboration2022-yt}. Additionally we also marginalize over the amplitude of the lens magnification parameters, with a wide prior, that are held fixed in the analysis presented in \cite{DES_Collaboration2022-yt}. As we show in Fig.~\ref{fig:resid_DV}, both galaxy-galaxy lensing and cross tomographic bin correlations of galaxy clustering are sensitive to the galaxy bias and lens magnification evolution. Therefore in Fig.~\ref{fig:cosmo_const} we analyze different probe combinations. We find approximately a $1\sigma$ bias in both $\Omega_m - S_8$ and $\Omega_m - w$ plane for both $\Lambda$CDM and $w$CDM cosmologies. We find that adding in CMB lensing cross-correlations results in more precise constraints but with similar biases.


However, as shown in Fig.~\ref{fig:Xinfer}, the de-correlation between galaxy clustering and galaxy-galaxy lensing becomes worse at higher redshifts. Therefore, cosmological analyses which constrain the evolution of growth with redshift (such as $\sigma_8(z)$), that are exclusively sensitive to the higher redshift bins can lead to up to 30\% biases depending upon the analysis configuration. Note that the constraints on $\sigma_8(z)$ even from current generation surveys at $z \leq 1$ are at better than 10\% level \citep{Abbott:2023:PhRvD:}.

\begin{figure*}
\includegraphics[width=\textwidth]{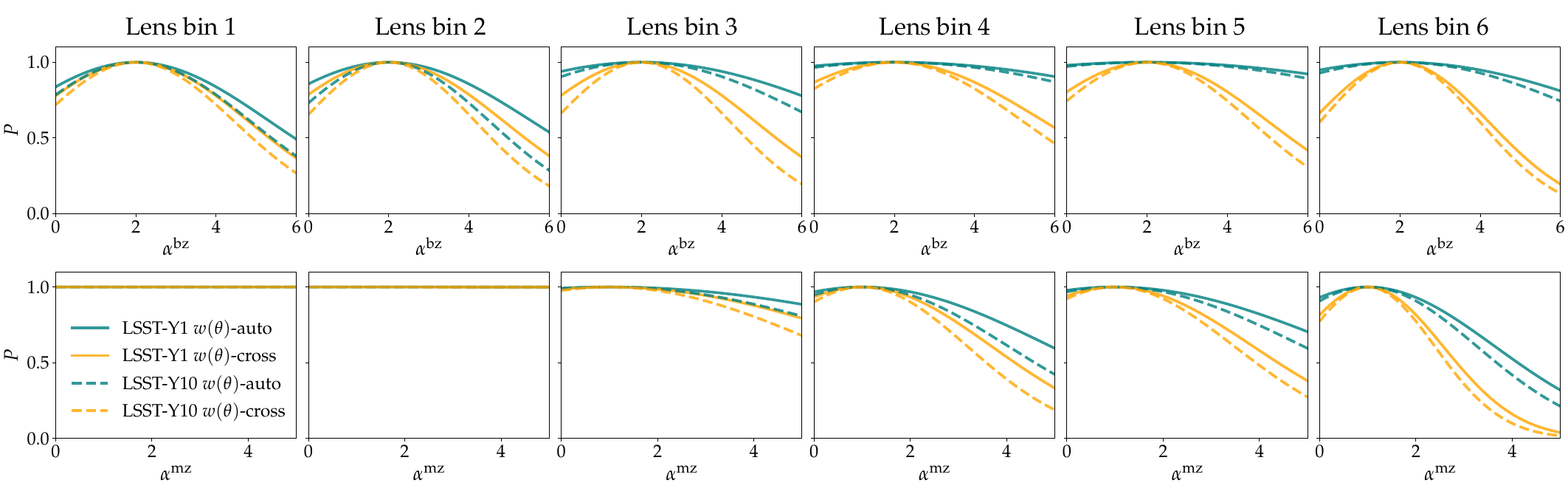}
\caption[]{
Fisher forecast of the constraints on the bias and lens magnification evolution parameters for the six  tomographic bins for various analysis configurations mentioned in the legend. In addition to the evolution parameters, this analysis marginalizes over all the parameters of $2\times 2$pt $\Lambda$CDM analysis presented in \cite{DES_Collaboration2021-gz} and lens magnification parameters.
}
\label{fig:bz_mz_constraints}
\end{figure*}

Since both the cross-tomographic bin galaxy clustering and galaxy-galaxy lensing measurements are impacted by the astrophysical evolution of galaxy bias and lens magnification. Therefore, a joint analysis of these two probes along with the auto-bin galaxy clustering measurements should constrain these evolution parameters. As the number of free parameters increase substantially when including the evolution parameters in the analysis which increases the nested sampling runtime substantially, we instead perform a Fisher forecast here. We show the Fisher forecast of these constraints assuming LSST Y1 and Y10 covariance in Fig.~\ref{fig:bz_mz_constraints}. Note that we marginalize over various cosmological as well as observational and modeling systematics parameters with same priors as described in \cite{DES_Collaboration2021-gz}. We find that using the cross-bin galaxy clustering measurements helps significantly in constraining these parameters. 

\section{Conclusions}
\label{sec:conclusion}

Cosmological analyses with imaging surveys of galaxies provide powerful measurements of the late time universe. The `2$\times $2pt' correlation functions include the clustering of galaxies and the galaxy-shear cross-correlation (galaxy-galaxy lensing). In this study we have quantified the impact of plausible levels of the evolution of galaxy bias (Fig.~\ref{fig:bz_sims}) and magnification on cosmological inferences. 

We find that galaxy clustering and galaxy-galaxy lensing can be sensitive to different effective redshifts depending upon the redshift distributions (Fig.~\ref{fig:zmean_probes}). Therefore, any evolution of galaxy bias and magnification  impacts the two probes differently. This general feature of 2$\times $2pt correlation functions leads to a stronger effect of bias evolution than the ``higher order'' contribution usually assumed. It is particularly important at high redshift, where  evolution is stronger, redshift distributions are broader, and the lensing kernel for source galaxies is steeper. (On the other hand, the galaxy-CMB lensing cross-correlations have broad lensing  kernels and are not affected nearly as much by evolution. )

Fig.~\ref{fig:chi2_values} quantifies the effect of bias and magnification evolution for galaxy-galaxy lensing measurements as well as cross-tomographic bin galaxy clustering correlations. We find that this redshift astrophysical evolution has a scale-independent impact on the datavectors (see Fig~\ref{fig:resid_DV}) that extends to the linear regime. This leads to an apparent de-correlation of the  amplitude of galaxy-galaxy auto-correlations and galaxy-galaxy lensing at the 10-15\% level, especially for the higher redshift bins (Fig.~\ref{fig:Xinfer}). This in turn leads to biased inferences if evolution is not part of the model. In $\Lambda$CDM and $w$CDM analysis, we find biases up to $1\sigma$ in the cosmological parameters of interest for both Y1 and Y10 survey specifications. Moreover, these biases would be especially important for extensions to the $\Lambda$CDM cosmological model that rely on higher redshifts, such as probing evolution of growth with redshifts \citep{DES_Collaboration2022-yt}.

We have checked whether bias and/or magnification evolution could explain two apparent anomalies discussed in the literature: ``lensing is low''\citep{Leauthaud2016-wk, Lange2018-fe, Amon2022-wu} and the detection of the decorrelation parameter $X_{\rm lens}$ \citep{Porredon2021-dl, Pandey2021-jc}. We find that the magnitude of the effects found in the above papers are probably not due to plausible levels of evolution though we have not carried out a detailed examination. Specifically, we find that including the evolution parameters in the fiducial $\Lambda$CDM analysis does not significantly improve the goodness of fit for the $2\times 2$pt and $3\times 2$pt analysis of \cite{DES_Collaboration2021-gz} when analyzing all six tomographic bins using the MagLim galaxy sample. This points to a different origin of the tension compared to the bias and magnification evolution. We leave a detailed analysis and interpreting the constraints on the evolution parameters to a future study.

We also quantify how changing the width of the redshift distribution of lens galaxies would change the impact of the astrophysical evolution. We find that a sample with narrower redshift distribution, for example obtained using red galaxy samples, would significantly reduce the biases caused by the evolution. We find that a joint analysis of galaxy clustering auto and cross-correlation, galaxy-galaxy lensing, and CMB lensing, including models for the evolution of galaxy bias and lens magnification, can provide tight and robust constraints on both cosmology and those astrophysical evolution models.

Our findings have the following implications for analysis choices for imaging surveys. 
\begin{itemize}
\item  In defining the galaxy sample, estimating the degree of bias evolution is important. Compromising on other desirable features, such as higher galaxy number density, may be worthwhile if the sample can be chosen to have slower, smoother or predictable evolution. Planning on a parallel HOD analysis or other ways to learn about bias evolution is valuable. 
\item Extra attention needs to be given to the width of redshift distributions when binning lens galaxy samples. Reducing this width can greatly mitigate the impact of bias or magnification evolution. 
\item Since our results suggest a stronger impact of bias (and magnification) evolution on cosmological analysis than previously understood, the model complexity (number of parameters) required to characterize bias evolution will be larger. E.g. at least one additional bias evolution parameter per $z$ bin may be needed for DES Y6 and possibly more for LSST Y1.  This will likely degrade constraints on cosmological parameters and lead to additional degeneracies: this requires further investigation. 
\item Cross-correlations of galaxy clustering across redshift bins can provide useful constraints on bias evolution and  should be part of the fiducial analysis. 
\item Finally, bias evolution will automatically be measured/constrained by such 2$\times$2pt or 3$\times$2pt analyses (this will be even more effective if CMB lensing is added to the datavector). Varying the evolution parameters can then provide a useful measurement of bias evolution across desired redshift ranges. 
\end{itemize}

\acknowledgments
\noindent This paper has undergone internal review by the LSST Dark Energy Science Collaboration (DESC). We kindly thank the internal reviewers Jonathan Blazek, Elisa Chisari and Constance Mahony for providing valuable comments, which helped us improve the quality and clarity of the paper. We also thank the Centro de Ciencias de Benasque Pedro Pascual for their hospitality while some of this work was carried out. 

\noindent The contributions from the primary authors are as follows: SP: Co-designed the project, lead the analysis and writing of the paper. CS: Co-designed the project, interpretation of the results and writing of the paper.  BJ: Contributed to the design of the experiment, help with interpretation of results, contributed to the paper text.

\noindent SP is supported by the Simons Collaboration on Learning the Universe. CS acknowledges financial support from Ramón y Cajal project RYC2021-031194-I and project PID2022-138123NA-I00, funded by MCIN/AEI/ 10.13039/501100011033 and by the “European Union NextGenerationEU/PRTR”.

\noindent The DESC acknowledges ongoing support from the Institut National de Physique Nucl\'eaire et de Physique des Particules in France; the Science \& Technology Facilities Council in the United Kingdom; and the Department of Energy, the National Science Foundation, and the LSST Corporation in the United States.  DESC uses resources of the IN2P3 Computing Center (CC-IN2P3--Lyon/Villeurbanne - France) funded by the Centre National de la Recherche Scientifique; the National Energy Research Scientific Computing Center, a DOE Office of Science User Facility supported by the Office of Science of the U.S.\ Department of Energy under Contract No.\ DE-AC02-05CH11231; STFC DiRAC HPC Facilities, funded by UK BEIS National E-infrastructure capital grants; and the UK particle physics grid, supported by the GridPP Collaboration.  This work was performed in part under DOE Contract DE-AC02-76SF00515.

\noindent This work made use of the following software packages: \texttt{astropy}\footnote{\url{https://www.astropy.org/}.}, \texttt{matplotlib}\footnote{\url{https://matplotlib.org/}.}, \texttt{numpy}\footnote{\url{https://numpy.org/}.} and \texttt{scipy}\footnote{\url{https://scipy.org/}.}.

\clearpage


\bibliographystyle{JHEP} 
\bibliography{des_y3kp}

\end{document}